\documentstyle[draft]{mn}
\title[Chemical evolution models]
      {Chemical evolution models with a new stellar nucleosynthesis}
\author[A. Giovagnoli and M. Tosi]
      {A. Giovagnoli$^{1}$ and M. Tosi $^{2}$\\
      $^{1}$Dipartimento di Astronomia, Universit\`a di Bologna,
       via Zamboni 33, 40126 Bologna, Italy;\\
      $^{2}$Osservatorio Astronomico, via Zamboni 33, 40126 Bologna, Italy}
\date{Received \hskip 1truecm     ; in original form 1993 June 8}
\pagerange{\pageref{firstpage}--\pageref{lastpage}}

\begin{document}
\label{firstpage}

\maketitle

\begin{abstract}
Numerical models for the chemical evolution of the Galaxy have been computed
with the new stellar yields published by Maeder (1992).
These metallicity dependent yields represent an important
improvement in the chemical evolution of galaxies but there are still
uncertainties in the stellar evolution which prevent completely
satisfactory results.
{}From the comparison
of the model predictions with the corresponding observational constraints
we find that Maeder's nucleosynthesis reproduces the oxygen and carbon
abundances and provides consistent $\Delta$Y/$\Delta$(O/H) ratios if a
significant amount of gas is accreted by the galactic disc during its whole
lifetime.
The lower mass limit for the black hole
formation (M$_{bh}$) must be larger than 22.5 M$_{\odot}$ to avoid oxygen
underproduction.
\par\noindent

\end{abstract}


\section{Introduction}

One of the major factors affecting the chemical evolution of galaxies is the
production (or destruction) of the elements inside the stars. The successive
pollution of the interstellar medium (ISM) due to the element ejection from
their parent star, weighted with the proper stellar initial mass function (IMF)
and coupled with the proper star formation rate (SFR), is in fact what produces
the chemical abundances currently observed in any galaxy. It has actually
been shown (Tosi 1988, Matteucci $\&$ Fran\c cois 1989)
that the abundance ratios between
different elements (e.g. N/O and O/Fe) or isotopes (e.g. $^{12}$C/$^{13}$C)
depend mostly on the stellar nucleosynthesis and very little on galactic
parameters like the star formation and infall rates if the accreted gas has
chemical abundances in roughly solar proportions.
Under these circumstances, the results of the computations on stellar evolution
and nucleosynthesis are of primary importance to model the chemical evolution
of galaxies, since they provide the amount of each element synthesized by
stars of any initial mass and the relative time taken by each star to eject it
in the ISM.
\par
In the last fifteen years several important works (e.g. Arnett 1978, Chiosi
$\&$ Caimmi 1979, Renzini $\&$ Voli 1981, Maeder 1981 and 1983, Woosley $\&$
Weaver 1986) have been devoted to stellar evolution models with detailed
nucleosynthesis calculations of the most diffuse elements ($^{4}$He, $^{12}$C,
$^{14}$N, $^{16}$O, etc). These studies have provided a grid of chemical yields
generally considered reliable enough to be adopted in galactic evolution
models.
Despite the well known uncertainties on several stellar parameters (e.g.
mixing length, mass loss, opacities) these yields have in fact proved roughly
consistent with each other and with the corresponding observational
constraints.
Most of these stellar models, however, were computed only for the solar
metallicity Z$_{\odot}$=0.02, thus leaving room for some concern about the
possibility of adopting their resulting yields also in epochs or in regions
where the ambient metallicity is much different from Z$_{\odot}$.
\par
Maeder (1992, hereinafter M92) has recently examined this problem and presented
the results of
nucleosynthesis in stars with various initial metallicities. In addition, his
stellar models are based on the most recent assumptions for all the input
physics (i.e. overshooting from convective cores, opacities and mass loss).
It is therefore of great interest to check what are the predictions of chemical
evolution models adopting these new yields (see also Carigi 1994 and Prantzos
1994 for the evolution of the solar neighbourhood). To this aim, we have
computed several numerical models of chemical evolution of the galactic disc
assuming M92 new yields, and compared their results with those obtained with
the models assuming the yields presented by previous
authors and described in the next section.
\par
In the next section we describe the model assumptions, indicating the major
differences between the two sets of adopted yields. In Section 3 we compare
the model predictions with the corresponding empirical data and in Section 4
we discuss the inferred conclusions.

\section{The Models}

The chemical evolution code employed here is described in detail in
Tosi $\&$ D{\'\i}az (1985) and Tosi (1988). It accounts for the stellar
lifetimes (i.e., avoids the instantaneous recycling approximation) and
follows the stellar nucleosynthesis of several elements.  The
galactic disc is divided into concentric rings; radial gas flows between
rings as well as infall of external gas onto each ring are allowed for.  The
star formation and the gas infall rates can vary with time and with
galactocentric distance. The adopted age of the disc is T=13 Gyr (e.g. Twarog
1980).
\par
The chemical evolution models have been computed with two alternative
assumptions for the stellar nucleosynthesis: the {\it standard} values defined
below and Maeder's (1992) new results. In the latter case, to follow all M92
prescriptions also the stellar lifetimes, remnants and returned fractions
have been changed accordingly to his new values. The nucleosynthesis results
are given in tables 5 and 6 of M92 for two initial stellar
metallicities, Z=0.001 and Z=0.02. We have then adopted the low metallicity
yields from the disc formation up to the time when the ISM reaches the
solar metallicity (an epoch different for different galactocentric distances)
and the Z=0.02 values afterwards. The {\it standard} yields, instead, are
complete only for solar metallicity models and we have therefore assumed their
constant contribution throughout the disc lifetime.

\subsection{Standard Yields}

Hereinafter we will refer to the yields adopted by D{\'\i}az $\&$ Tosi (1986,
hereinafter DT86) and
Tosi (1988) as the {\it standard} yields. These values are derived from the
nucleosynthesis computations by Renzini $\&$ Voli (1981) for low and
intermediate mass stars and by Arnett (1978), revisited by Chiosi $\&$ Caimmi
(1979), for massive stars. To take into account also the effect of massive
star winds, the latter yields are combined with Maeder's (1981, 1983) values.
On this basis, the elements for which the evolution can be modeled are
$^{4}$He, $^{12}$C, $^{13}$C, $^{14}$N and $^{16}$O. We recall that Renzini
$\&$ Voli showed that during the envelope burning on the asymptotic giant
branch of intermediate mass stars some primary $^{13}$C and $^{14}$N are
produced. Thus, none of these five elements can be considered completely
secondary (i.e. requiring a previous generation of stars to be synthesized).
To resume the overall results of this {\it standard} nucleosynthesis: $^{4}$He
is produced by stars of any initial mass, $^{12}$C by both high and
intermediate
mass stars, $^{13}$C and $^{14}$N by intermediate mass stars, and
$^{16}$O by massive stars only.
\par
As already discussed by DT86 and Tosi (1988), these yields
suffer of several problems. For instance, the chemical evolution model
presented below assume a mixing length parameter $\alpha$=1.5 in the
stellar envelope (see Renzini $\&$ Voli 1981) and no overshooting from
convective cores (i.e. $\lambda$=0) whereas, according to more recent
views on this field, a more proper combination of these
two parameters could be with a slightly smaller value for $\alpha$ or
a slightly larger value for $\lambda$. The uncertainties on these
values have relevant effects on the determination of the correct element
abundances. In fact,
a smaller $\alpha$ implies less nitrogen, due to a lower production of
its primary component during the envelope burning in asymptotic giant branch
stars, while a larger $\lambda$ implies more oxygen, which is produced in
the cores of massive stars, and less nitrogen, because of the reduced size
of the stellar envelopes where $^{14}$N is mostly synthesized (e.g. Greggio
and Tosi 1986, but see also Serrano 1986). On the basis of the most recent
stellar evolution theories it is also possible that the contribution of
intermediate mass stars to the enrichment of $^{12}$C, $^{13}$C and $^{14}$N
has been slightly overestimated (Renzini, private communication).
\par
It is also worth mentioning that the $^{12}$C(${\alpha,\gamma}$)$^{16}$O
reaction rate may be faster than assumed in these {\it standard} models, thus
producing more oxygen and less carbon, and that the actual nucleosynthesis
of $^{13}$C should be quite different to allow a galactic radial gradient
of $^{12}$C/$^{13}$C as steep as that derived from molecular clouds
observations (Tosi 1988, D'Antona $\&$ Matteucci 1991).

\subsection{M92 New Yields}

The {\it standard} yields were all computed for stellar models with initial
solar composition.  To overcome this limitation and to introduce the proper
updating on several parameters,  Maeder has recently computed a complete
grid of stellar evolution tracks with different initial metallicity Z and
published (M92) the results of their nucleosynthesis for Z=0.001 and
Z=Z$_{\odot}$=0.02.
The major differences between his new models and the {\it standard} ones
are the inclusion of a moderate ($\lambda \simeq $0.2) overshooting
from convective cores, the adoption of new opacity tables (Roger $\&$
Iglesias 1992) and different evaluations of the stellar remnants after
the wind mass loss and of the limiting mass for black hole formation.
He also takes into account the dependence of the mass
loss rate on metallicity and correctly assumes different initial helium
content for different initial metallicity (Y=0.243 for Z=0.001 and
Y=0.30 for Z=0.02).
\par
The ranges of masses where $^4$He, $^{12}$C, $^{14}$N and
$^{16}$O are produced are the same as in the {\it standard} nucleosynthesis.
Since M92 stellar models for low and intermediate mass stars do not reach
the final evolutionary phases, the corresponding heavy element
contributions are not given in that paper and we take them from Renzini
$\&$ Voli (1981).
\par
An important conclusion of M92 is that the nucleosynthetic production
strongly depends on the stellar initial metallicity Z. In particular: for
increasing Z,
$^{16}$O is strongly depleted, $^{12}$C is highly enhanced in stars more
massive than 25 M$_{\odot}$ but fairly reduced in smaller stars, and
$^{4}$He is highly enhanced in very massive stars and roughly constant
in the others.
\par
If we compare the amount of each of these elements ejected by massive stars
in M92 models for Z=0.02 with the corresponding amount in
{\it standard} models (see also Figs 7 to 10 in M92),
we find that in M92 the $^{4}$He enrichment is larger,
$^{12}$C is much larger for very massive stars,
$^{16}$O is slightly larger for 10$\leq$M/M$_{\odot}\leq$25
and much smaller for more massive stars. For Z=0.001, M92's ejected masses are
similar to the {\it standard} ones with Z=0.02 for $^{4}$He, lower for $^{12}$C
and slightly larger for $^{16}$O. The total metallicity Z produced by massive
stars is fairly lower than the {\it standard} value for Z=0.02 and only
slightly lower for Z=0.001.
\begin{figure}
 \centering
 \vspace{7cm}
 \caption{Oxygen abundance distribution in the galactic disc as derived from
 Peimbert (1979) and Shaver's et al. (1983) observations of HII regions
 (dots with average
 error bars) and from model predictions. The thick solid line
 represents the {\it standard} model, the thin solid line model 1, the dotted
 line model 2, the short-dashed line model 3, the long-dashed line model 4,
 the short-dash-dotted line model 5, the long-dash-dotted line model 6.}
 \label{sample-figure}
\end{figure}
\section{Model results and observational data}

The predictions of chemical evolution models for our galactic disc based
on {\it standard} nucleosynthesis have been compared by Tosi (1988)
with the corresponding observational data.  Here we simply recall the results
more relevant to our current issue and show only the predictions of one of
the models in better agreement with all the observational constraints. From
now on this model will be referred to as the {\it standard} model; it assumes
an exponentially decreasing SFR with $e$-folding time $\tau$=15 Gyr
and initial value derived from the observed amount of current gas
and total mass in each ring, an
almost constant infall rate after the disc formation with uniform density
across the entire disc of $4 \cdot 10^{-3} M_{\odot} kpc^{-2} yr^{-1}$,
and Tinsley's (1980) IMF.
\par
The radial gradients of the nitrogen and oxygen abundances derived from
HII region observations (Peimbert 1979, Shaver et al. 1983) are very well
reproduced
by the {\it standard} model. As for the absolute values of their abundances,
the predicted oxygen content is in good agreement with the data (Fig. 1,
where the {\it standard} model is represented by the thick solid line), whereas
a value of the mixing length parameter $\alpha$ intermediate between those
available in the literature ($\alpha$=1.0 and $\alpha$=1.5) would be
required to achieve the same agreement for nitrogen. In fact, too
much $^{14}$N is produced
if $\alpha$=1.5, but its resulting abundance is too low if $\alpha$=1.0
(see DT86, fig.4). Similarly, an intermediate value of $\alpha$, say $\sim$
1.2, would allow to reproduce both the trend and the absolute values of the
N/C ratio with C/H as derived by Laird (1985) from the spectroscopy of more
than a hundred stars in the solar neighbourhood (see DT86, fig.7).

\begin{figure}
 \centering
 \vspace{7cm}
 \caption{AMR in the solar neighbourhood. The dots with error bars represent
Twarog's (1980) data, the model symbols are as in Fig.1. Note that [Fe/H]
is not the iron abundance but represents the global metallicity normalized
to its solar value.}
 \label{sample-figure}
\end{figure}

\par
The predicted distribution with time of the overall metallicity in the
solar ring is in agreement with Twarog's (1980) age-metallicity relation
(thick solid line in Fig.2)  except for the very early epochs after the disc
formation where the model predicts too low metallicities because it assumes an
initial Z=0.  We emphasize that Twarog's [Fe/H] is not the iron abundance but
stands for the global metallicity normalized to the solar value. We keep using
his traditional notation but warn the reader that the actual time-behaviour
of iron may be quite different from that shown in Fig.2.
\par
The assumption of primordial initial metallicity also leads
to a slight excess of low metallicity stars predicted by the {\it standard}
model in the comparison with the stellar frequency distribution with Z derived
by Pagel $\&$ Patchett (1975) from observations of G-dwarf stars in the solar
neighbourhood. The histogram corresponding to Pagel's (1989) updated version
of these data is shown in Fig.3a together with
the predictions of our {\it standard} model (thick solid line). Note that, for
a more sensible interpretation of these data, we have followed Pagel's
suggestion and plotted the stellar distribution as a function of the oxygen
abundance rather than the overall Z metallicity, since the latter is the sum
of elements produced in too different sites and at too different epochs.

To allow for an immediate comprehension of what is the effect of assuming
M92 stellar nucleosynthesis, the thin solid line in Figs 1 to 4
shows the predictions of a model (model 1) with all the  parameters
(namely: IMF, SFR and infall rate) identical to those of the {\it standard}
model. As already mentioned in the previous section, we adopt the Z=0.001
nucleosynthesis provided by M92 up to the epoch when the ISM reaches Z=0.02
and its solar nucleosynthesis afterwards.
\par
Model 1 reproduces well the observational properties of our Galaxy. At the
present epoch it predicts  oxygen abundances in the ISM of the whole disc
in good agreement with the observational data (HII regions) and with the
{\it standard} model (thin solid line in Fig. 1). The divergence of
the solid lines at the outermost ring is due to the differences between
the Z=0.001 and Z=0.02 oxygen yields.
The former are higher than the latter and since the outer ring reaches the
solar metallicity much later than the others, it is enriched for
longer times by the higher O-yields. For this reason it shows at the present
time higher oxygen abundances.
\par
The frequency distribution with [O/H] of the G-dwarfs in the solar
ring predicted by model 1 looks consistent with the data (thin solid line
in Fig.3a), although model 1 predicts too many G-dwarfs with low oxygen.

\begin{figure}
 \centering
 \vspace{8cm}
 \caption{Fraction of G-dwarf stars in the solar neighbourhood as a
function of their oxygen abundance. The histogram corresponds to Pagel's (1989)
observational data; the model symbols are as in Fig.1.}
 \label{sample-figure}
\end{figure}

\par
Fig.4 shows the model results for the radial distribution of $^{12}$C.
These predictions can be compared with the corresponding observational data
available for the solar neighbourhood.
The vertical line in Fig.4 represents the range of carbon abundances derived
by Laird (1985) from the spectroscopy of 116 nearby stars and its length
includes both his quoted observational error ($\pm$0.25 dex) and the abundance
spread probably due to the different ages and/or initial metallicity of the
examined stars. As usual, the thick solid curve in this figure represents the
predictions of the {\it standard} model, whereas the thin solid line
corresponds to the analogous model 1 with M92 nucleosynthesis.
The {\it standard} predicted abundances fall within the observed range as
well as those based on M92 but the latter show a rapid decrease in the outer
regions of the disc. Since the carbon contribution from low and intermediate
mass stars is the same as that of the {\it standard} model (i.e. that from
Renzini $\&$ Voli 1981), we infer that the higher abundances at inner radii
and the rapid decrease in the outer ones are due to the yields of M92 massive
stars that produce more carbon at Z=Z$_{\odot}$.
Despite their larger values, we find that the carbon abundances predicted by
the combination of Maeder's massive stars yields with those from intermediate
mass stars are consistent with the observed data contrary to what is
suggested by Prantzos et al. (1994).

\par
The age-metallicity relation predicted by model 1 is roughly
consistent with the corresponding observational data (Fig. 2) and does not
differ significantly from the predictions of the {\it standard} model.

\par
\begin{figure}
 \centering
 \vspace{7cm}
 \caption{Carbon abundance distribution in the galactic disc. The
vertical line corresponds to the range of values derived by Laird (1985)
from observations of stars in the solar neighbourhood. The model symbols
are as in Fig.1.}
\end{figure}

We have computed other numerical models with various choices of the
evolutionary parameters SFR, IMF and infall, and covering as much as possible
the range of reasonable values of these parameters.
Table 1 lists the most significant of these models, which are described in
this paper.
\par
One of the most controversial issue in  galaxy evolution concerns the
presence of infall of gas on spiral galaxies. The only observational
evidence of infall are High Velocity Clouds (HVC's) and mostly Very High
Velocity Cloud's. In order to check the role of infall in the
chemical evolution based on M92 yields we have assumed no accretion after
the disc formation (model 2), thus removing any dilution of the ISM.
The dotted line in Fig.1
shows that the oxygen predictions of model 2 do not reproduce
the observational data, even if the oxygen depletion with increasing Z
makes the disagreement less dramatic than with standard nucleosynthesis.
As often found for no infall model, the slope
of the abundance gradient is not reproduced either. Fig. 4 shows that also the
carbon predicted abundance is inconsistent
with the observational constraint in the solar neighbourhood.
As for the metallicity distribution of the G-dwarfs (Fig.3a), this closed box
model predicts as usual more stars with low [O/H].
This is because the disc does not accrete mass during its
life and starts forming stars already with its final mass. Thus, a larger
number of stars are formed at early epochs, and therefore with low initial
metallicity, than in models with infall. For the same reason, the
age-metallicity relation (dotted line in Fig.2) shows lower abundances at
early epochs. It also shows too large abundances at recent epochs, due to
the lack of any dilution of the ISM enrichment.
\begin{table}
\centering
\caption{Models with M92 nucleosynthesis }
\medskip
\begin{tabular} {lccc}
\hline
\bigskip
\label{symbols}
 Model  & SFR  & Infall  & IMF   \\
 number    &                & ($M_{\odot} kpc^{-2} yr^{-1}$)   &         \\
           &                &                                  &         \\
 $~~1 $     & $\tau=15~Gyr$  & $B=4\cdot 10^{-3}$   & Tinsley  \\
 $~~2 $     & $\tau=15~Gyr$  & $B=~0 $                & Tinsley  \\
 $~~3 $     & $\tau=15~Gyr $ & $B=4\cdot 10^{-3}$   & Salpeter \\
 $~~4 $     & $\propto g^n~\left( n=1\right)$ & $B=2\cdot 10^{-3}$ & Tinsley
\\
 $~~5^{\dagger}$     & $\tau=15~Gyr$  & $B=4\cdot 10^{-3}$   & Tinsley \\
 $~~6^{\dagger}$     & $\tau=15~Gyr$  & $B=4\cdot 10^{-3}$ & Salpeter \\
              &                &                        &         \\
\hline
\end{tabular}
$^{\dagger}$Models 1 and 5 differ only in the black hole mass limit. The same
applies to models 3 and 6.
\end{table}
It is then clear that removing infall worsens the predicted stellar
metallicity distribution. Intermediate values of infall obviously provide
intermediate results and we therefore conclude that even with metallicity
dependent yields the amount of accreted gas must be around $B=4\cdot 10^{-3}
M_\odot kpc^{-2} yr^{-1}$ as found with constant yields.

\par
In model 3 we have adopted
Salpeter's IMF instead of Tinsley's. Salpeter's mass function assumes
a larger fraction of massive stars because the exponent for that
mass range is -2.35 instead of Tinsley's -3.3, and this
increases the amount of oxygen available at the end of the Galaxy
evolution, since this element is produced only by high
mass stars. Model 3 (short-dashed line in all figures) assumes an infall
rate of $4\cdot 10^{-3} M_{\odot} kpc^{-2} yr^{-1}$ and the only difference
with model 1 resides in the IMF.
The oxygen distribution predicted by model 3 is shown in Fig. 1 where the
oxygen
overproduction due to the larger fraction of massive stars is apparent. The
age-metallicity relation (Fig. 2) is in agreement with
the observational constraints while the carbon predicted distribution (Fig.4)
is completely out of the data range. This inconsistency is due to the larger
carbon production by massive stars with Z=0.02 combined with Salpeter's larger
fraction of massive stars. Model 3
predicts many G-dwarfs
(Fig. 3b) with ``solar" oxygen abundance because it produces oxygen
faster than model 1 and therefore reaches the solar abundance earlier.
\par
Models with the star formation simply proportional
to the gas density show by definition a radial distribution of the current SFR
equal to that of the observed gas density and therefore too flat with respect
to that derived from observations of recently formed objects (see e.g. Lacey
$\&$ Fall 1985, Tosi 1988). Nonetheless, since this parametrization is still
fairly popular we present their predictions for sake of completeness.
A typical example of this class is model 4 (long-dashed lines in
all figures) which assumes Tinsley's IMF and an infall rate of $2\cdot 10^{-3}
M_{\odot} kpc^{-2} yr^{-1}$, a value quite lower than that required by the
models discussed above with a different star formation law. This is because,
in this case, the effect of infall is not simply to dilute the ISM abundances
but also to increase the region gas density thus enhancing the SFR and the
ISM chemical enrichment. Due to the different balancing between SFR and
infall, the predicted oxygen (Fig.1) and carbon (Fig.4) abundances are fairly
larger than with model 1. In the case of oxygen the curve predicted by model 4
still lies well within the observed range of abundances, but for carbon it is
at the upper edge of the data. The large star
formation activity occurred in this model at early epochs when most of the gas
was available generates in the
solar ring a large number of G-dwarfs with low oxygen content, and the low
activity of more recent epochs, when less gas is available, generates only
few stars with {\it solar} oxygen. These features make the G-dwarf
distribution inconsistent with the empirical histogram of Fig.3b.
\par
Finally, we have examined the effect of assuming an upper mass limit for the
stellar contribution to the ISM, $M_{bh}$.
All the stars with initial mass higher than $M_{bh}$
collapse into a black hole and do not contribute to the ISM enrichment
with a final explosion.
In M92 various limiting masses $M_{bh}$ for black hole formation are
considered. Considering Maeder's arguments and bearing in mind that
the supernova (SN 1987A) recently exploded in the Large Magellanic Cloud
originated from a $20 M_{\odot}$ star, the lowest reasonable choice for
$M_{bh}$ seems be $22.5 M_{\odot}$ (M92 case c).
In Figs 1 and 4 (short-dash-dotted line) we present the corresponding
effect on the predictions of model 1. From these two figures one can see that
the decrease in oxygen is much stronger than in carbon because of the different
mass range where these elements are mostly produced. This choice of $M_{bh}$
cuts significantly
the range of masses contributing to the oxygen enrichment and leads to current
abundances totally inconsistent with the data.
The age-metallicity relation is not affected by the $M_{bh}$ introduction
whereas the G-dwarf distribution (Fig. 3a) predicts too many stars at low
oxygen abundances and it does not reproduce the "solar" observational value.
It is then unacceptable, as a priori obvious and already found by Maeder.
\par
As further check we have applied the same limit $M_{bh}$= 22.5 M$_{\odot}$
to model 3 which was found to predict too large oxygen abundances because of
the large fraction of massive stars.
Model 6 (long-dash-dotted line in all figures) has then been calculated as
model 3 but with the introduction of the mass limit for black hole formation.
As expected the oxygen abundance is sensibly depleted and is now roughly
consistent with the observational
data. Carbon however is not reduced enough and remains well above the
observation range. The G-dwarfs distribution (Fig. 3b) is not satisfactory
either.
\section{Discussion}
In the previous section we have compared the model predictions on the C and O
abundances based on Maeder's (1992) nucleosynthesis with the corresponding
values observed in the Galaxy.
It is clear that the strong metallicity dependence of M92 nucleosynthesis
affects the chemical evolution results. For instances, models 1, 4 and 5 show
a particular behaviour in the outer regions of the disc where the lower
metallicity yields are used for most of the galactic life.
On the other hand some of the general results derived (e.g. Tosi 1988)
with the classical nucleosynthetic prescriptions are still valid with M92
yields.
Models without infall (e.g. model 2) must be rejected because they
do not allow to reproduce either the G-dwarfs distribution or
the O and C abundances.
Different choices of SFR (model 4) and IMF (model 3) generally worsen the
agreement between theoretical predictons and observational constraints.
\par
An interesting effect of metallicity dependent yields is on the
helium-metallicity relation $\Delta$Y/$\Delta$Z which has important
cosmological implications since it provides, by extrapolation to Z=0, the
primordial value of $^4$He.
Up to date, there are no direct estimates of the true
$\Delta$Y/$\Delta$Z. What is actually observed is not the global metallicity
but oxygen (mostly in galactic and extragalactic HII regions) and the
empirical $\Delta$Y/$\Delta$Z is derived from its abundances assuming a
linear correlation between Z and O.
This correlation is extremely controversial: first of all because, as
shown by M92 models, it is not actually linear, and second because different
authors assume rather different slopes. To overcome these large uncertainties,
and for a correct comparison of the model predictions with the corresponding
observational data, we have chosen to examine the $\Delta$Y/$\Delta$(O/H)
resulting from the models because the empirical estimates of this ratio are
directly derived from observations. The most updated and reliable value is
provided by Pagel's et al. (1992) from a large sample of H II regions in spiral
and irregular galaxies, $\Delta$Y/$\Delta$(O/H) = 125 $\pm$ 40. The values
predicted by our chemical evolution models based on {\it standard} and on
M92 nucleosynthesis are 63 and $\sim$72, respectively. If we take into
account the fact that the observational data refer mostly to metal poor
galaxies and that some authors (e.g. Campbell 1992, Lenzuni $\&$ Panagia 1993)
suggest that the ratio is lower in high metallicity objects (but see also
Pagel 1993), we see that the {\it standard} and M92 values may both be roughly
consistent with the empirical ratio. It is worth noticing that using the true
Z abundance M92 finds $\Delta$Y/$\Delta$Z = 3 - 6 and out {\it standard} model
gives $\Delta$Y/$\Delta$Z = 3 which is the ratio traditionally derived from
observatons of spiral galaxies (e.g Lequeux et al. 1979, Torres-Peimbert,
Peimbert \& Fierro 1989 and references therein).
\par
Following M92's suggestion, we have also checked the effect on the chemical
evolution of the galactic disc of assuming an upper limit to the mass of the
stars contributing to the ISM enrichement. We confirm that this limit cannot
be very low (M$_{bh}$= 22.5 M$_{\odot}$ is already too low) as already
suggested by M92 and Prantzos 1994.
\par
We believe that for a more detailed analysis of the mass limit in better
agreement with the observed chemical features of the Galaxy, we should
wait for update calculations of the stellar nucleosynthesis not only of massive
stars. In fact, the major observational constraints involve elements like
helium, carbon, nitrogen and iron which are mostly produced by stars smaller
than those examined by M92. A revision of the classical work by Renzini and
Voli (1981) with the most updated input physics would be necessary to achieve
more significant results.
\medskip
\bigskip
\\
{\bf Acknowledgements}
We warmly thank Laura Greggio, Alvio Renzini and Claudio Ritossa for useful
conversations on the stellar evolution models and Francesca Matteucci for
interesting comments and discussions. We are grateful to Andr\'e Maeder for
a nice and lively discussion which has helped us to better understand the
problem and to significantly improve the paper.

\bsp

\label{lastpage}

 \end{document}